\documentclass[a4paper,fleqn]{cas-dc}

\usepackage[numbers]{natbib}

\def\tsc#1{\csdef{#1}{\textsc{\lowercase{#1}}\xspace}}
\tsc{WGM}
\tsc{QE}
\tsc{EP}
\tsc{PMS}
\tsc{BEC}
\tsc{DE}

\newcommand{\para}[1]{\vspace{2pt}\noindent\textbf{#1.~}}
\newcommand{\ignore}[1]{}
\newcommand{\system}{\sloppy{ParityFuzz\@}}

\usepackage{listings, xcolor}
\usepackage{pifont}
\usepackage{ulem}
\usepackage{array,framed}
\usepackage{setspace}
\usepackage{latexsym,fancyhdr,url}

\usepackage{enumerate}
\usepackage{xspace}
\usepackage{multirow}
\usepackage{microtype}

\usepackage{csvsimple}
\usepackage{colortbl}
\usepackage{listings}
\usepackage{booktabs}
\usepackage{multirow}
\usepackage{hyperref}
\usepackage{tcolorbox}
\usepackage{float}
\usepackage{adjustbox}
\usepackage{bigstrut}
\usepackage{tabularx}
\usepackage{pdfpages}      
\usepackage{hyperref}      
\usepackage{graphicx}
\usepackage{subcaption}
\usepackage{amsmath}
\usepackage{amssymb}
\usepackage{url}
\usepackage{hyperref}

\usepackage{template/lstlinebgrd}
\definecolor{cgreen}{RGB}{106,153,85}
\definecolor{hred}{RGB}{255,204,204}

\definecolor{NGreen}{RGB}{0,176,80}
\definecolor{verylightgray}{rgb}{.97,.97,.97}
\definecolor{githubgrey}{RGB}{124, 135, 142}
\definecolor{githubgreen}{RGB}{230, 255, 236}
\definecolor{githubpink}{RGB}{255, 235, 233}
\lstdefinelanguage{Solidity}{
  keywords=[1]{anonymous, assembly, assert, balance, break, call, callcode, case, catch, class, constant, continue, constructor, contract, debugger, default, delegatecall, delete, do, else, emit, event, experimental, export, external, false, finally, for, function, gas, if, implements, import, in, indexed, instanceof, interface, internal, is, length, library, log0, log1, log2, log3, log4, memory, modifier, new, payable, pragma, private, protected, public, pure, push, require, return, returns, revert, selfdestruct, send, solidity, storage, struct, suicide, super, switch, then, this, throw, transfer, true, try, typeof, using, value, view, while, with, addmod, ecrecover, keccak256, mulmod, ripemd160, sha256, sha3}, 
  keywordstyle=[1]\color{blue}\bfseries,
  keywords=[2]{address, bool, byte, bytes, bytes1, bytes2, bytes3, bytes4, bytes5, bytes6, bytes7, bytes8, bytes9, bytes10, bytes11, bytes12, bytes13, bytes14, bytes15, bytes16, bytes17, bytes18, bytes19, bytes20, bytes21, bytes22, bytes23, bytes24, bytes25, bytes26, bytes27, bytes28, bytes29, bytes30, bytes31, bytes32, enum, int, int8, int16, int24, int32, int40, int48, int56, int64, int72, int80, int88, int96, int104, int112, int120, int128, int136, int144, int152, int160, int168, int176, int184, int192, int200, int208, int216, int224, int232, int240, int248, int256, mapping, string, uint, uint8, uint16, uint24, uint32, uint40, uint48, uint56, uint64, uint72, uint80, uint88, uint96, uint104, uint112, uint120, uint128, uint136, uint144, uint152, uint160, uint168, uint176, uint184, uint192, uint200, uint208, uint216, uint224, uint232, uint240, uint248, uint256, var, void, ether, finney, szabo, wei, days, hours, minutes, seconds, weeks, years},  
  keywordstyle=[2]\color{teal}\bfseries,
  keywords=[3]{block, blockhash, coinbase, difficulty, gaslimit, number, timestamp, msg, data, gas, sender, sig, value, now, tx, gasprice, origin},  
  keywordstyle=[3]\color{violet}\bfseries,
  identifierstyle=\color{black},
  sensitive=false,
  comment=[l]{//},
  morecomment=[s]{/*}{*/},
  commentstyle=\color{githubgrey}\ttfamily,
  stringstyle=\color{red}\ttfamily,
  morestring=[b]',
  morestring=[b]",
}
\lstdefinelanguage{Log}{
  morekeywords={Warning, CHC},
  keywordstyle=\color{red}\bfseries,
  morecomment=[l]{-->},
  commentstyle=\color{gray},
  morecomment=[l]{|},
  stringstyle=\color{blue},
  sensitive=true,
  moredelim=[is][\color{red}]{-->}{},
  moredelim=[is][\color{red}]{|}{},
  moredelim=[il][\color{red}]{^},
}

\lstdefinelanguage{Solidity}{
  keywords=[1]{anonymous, assembly, assert, balance, break, call, callcode, case, catch, class, constant, continue, constructor, contract, debugger, default, delegatecall, delete, do, else, emit, event, experimental, export, external, false, finally, for, function, gas, if, implements, import, in, indexed, instanceof, interface, internal, is, length, library, log0, log1, log2, log3, log4, memory, modifier, new, payable, pragma, private, protected, public, pure, push, require, return, returns, revert, selfdestruct, send, solidity, storage, struct, suicide, super, switch, then, this, throw, transfer, true, try, typeof, using, value, view, while, with, addmod, ecrecover, keccak256, mulmod, ripemd160, sha256, sha3}, 
  keywordstyle=[1]\color{blue}\bfseries,
  keywords=[2]{address, bool, byte, bytes, bytes1, bytes2, bytes3, bytes4, bytes5, bytes6, bytes7, bytes8, bytes9, bytes10, bytes11, bytes12, bytes13, bytes14, bytes15, bytes16, bytes17, bytes18, bytes19, bytes20, bytes21, bytes22, bytes23, bytes24, bytes25, bytes26, bytes27, bytes28, bytes29, bytes30, bytes31, bytes32, enum, int, int8, int16, int24, int32, int40, int48, int56, int64, int72, int80, int88, int96, int104, int112, int120, int128, int136, int144, int152, int160, int168, int176, int184, int192, int200, int208, int216, int224, int232, int240, int248, int256, mapping, string, uint, uint8, uint16, uint24, uint32, uint40, uint48, uint56, uint64, uint72, uint80, uint88, uint96, uint104, uint112, uint120, uint128, uint136, uint144, uint152, uint160, uint168, uint176, uint184, uint192, uint200, uint208, uint216, uint224, uint232, uint240, uint248, uint256, var, void, ether, finney, szabo, wei, days, hours, minutes, seconds, weeks, years},  
  keywordstyle=[2]\color{teal}\bfseries,
  keywords=[3]{block, blockhash, coinbase, difficulty, gaslimit, number, timestamp, msg, gas, sender, sig, value, now, tx, gasprice, origin},  
  keywordstyle=[3]\color{violet}\bfseries,
  identifierstyle=\color{black},
  sensitive=false,
  comment=[l]{//},
  morecomment=[s]{/*}{*/},
  commentstyle=\color{gray}\ttfamily,
  stringstyle=\color{black}\ttfamily,
  morestring=[b]',
  morestring=[b]"
}
\lstdefinelanguage{Rust}{
  keywords={fn, let, mut, if, else, match, while, for, in, loop, break, continue, return, struct, enum, impl, trait, const, static, use, pub, mod, crate, super, Self, self, ref, move, as, async, await, dyn, unsafe},
  keywordstyle=\color{blue}\bfseries,
  ndkeywords={true,false,None,Some,Ok,Err},
  ndkeywordstyle=\color{orange}\bfseries,
  identifierstyle=\color{black},
  sensitive=true,
  comment=[l]{//},
  morecomment=[s]{/*}{*/},
  commentstyle=\color{gray}\ttfamily,
  stringstyle=\color{red}\ttfamily,
  morestring=[b]',
  morestring=[b]",
}

\begin{document}
\newcommand{\bugs}{64}
\newcommand{\confirmedBugs}{27}
\newcommand{\fixedBugs}{11}
\newcommand{\fixedByUsBugs}{11}

\let\WriteBookmarks\relax
\def\floatpagepagefraction{1}
\def\textpagefraction{.001}
\shortauthors{Bowei Su et~al.}

\title [mode = title]{\system{}: Finding Inconsistencies across Solidity Compilers via Fine-Grained Mutation and Differential Analysis}                      



\author[1]{Bowei Su}[style=chinese]
\ead{subw3@mail2.sysu.edu.cn}
\credit{Methodology, Validation, Writing – original draft}

\author[1]{Mingxi Ye}[style=chinese]
\cormark[1]
\ead{yemx6@mail2.sysu.edu.cn}
\credit{Methodology, Validation, Writing – review and editing}

\author[1]{Yuhong Nan}[style=chinese]
\ead{nanyh@mail.sysu.edu.cn}
\credit{Methodology, Resources, Writing – review and editing}

\author[1]{Peilin Zheng}[style=chinese]
\ead{zhengplin@mail.sysu.edu.cn}
\credit{Data curation, Software, Writing – review and editing}

\author[1]{Zibin Zheng}[style=chinese]
\ead{zhzibin@mail.sysu.edu.cn}
\credit{Investigation, Validation, Writing – review and editing}

\affiliation[1]{organization={School of Software Engineering, Sun Yat-sen University},country={Zhuhai, 519082, China}}






\cortext[cor1]{Corresponding author}


\begin{abstract}
The Solidity smart contract ecosystem has rapidly grown, leading to multiple compilers targeting different blockchain platforms or offering improved compilation efficiency. Although many compilers aim to be compatible with the primary Solidity compiler (Solc), significant inconsistencies in compilation and execution remain. These inconsistencies hinder contract migration, mislead developers during debugging, and may introduce exploitable vulnerabilities, causing potential financial losses. Existing testing techniques mainly focus on bugs within a single compiler or perform differential testing across compilers targeting the same environment. However, these approaches are inadequate for detecting inconsistencies across Solidity compilers, as they lack mechanisms to explore inconsistency-triggering conditions and do not support comparing bytecode generated for different environments.

To address this gap, we propose \system{}, a cross-compiler differential testing framework for Solidity. \system{} operates in three stages. First, it generates a rich set of mutation rules (i.e., syntax-oriented and boundary-oriented mutation rules) by analyzing source code (i.e., compiler and execution environment). Second, it employs a reinforcement learning-based strategy to select the most promising rules for mutating test programs. Finally, it detects inconsistencies by compiling and executing these programs on multiple compilers, then normalizing and comparing their results. Our evaluation demonstrates that \system{} is both efficient and effective. It improves test program generation, achieving up to an 18$\times$ higher compilation success rate and 1.8$\times$ greater code coverage compared to state-of-the-art fuzzers. In total, \system{} has uncovered \bugs{} previously unknown inconsistencies across six popular compilers. Notably, our findings have led to 11 fixes by developers and received a bounty from the Polkadot community.
\end{abstract}




\maketitle

\section{Introduction}

The Solidity smart contract ecosystem has flourished in recent years, leading to the emergence of multiple Solidity compilers. There are two main reasons for this trend. First, some blockchain platforms aim to attract developers by supporting Solidity. Since these platforms do not natively support the Ethereum Virtual Machine (EVM), they develop new compilers that enable the deployment of Solidity smart contracts. Second, additional compilers have been developed to improve the performance and optimization of Solidity compilers, even on platforms that already support EVM.

Several popular Solidity compilers, including Solc~\cite{solc}, Revive~\cite{revive}, Zksolc~\cite{zksolc}, Solang~\cite{solang}, Sold~\cite{sold}, and Solar~\cite{solar}, support blockchain ecosystems that collectively hold more than \$100 billion in Total Value Locked (TVL), according to DefiLlama~\cite{DefiLlama}. This highlights the critical role that Solidity compilers play in today’s blockchain infrastructure. However, the increasing number of Solidity compilers has also introduced significant inconsistencies among them.

These inconsistencies can have serious real-world consequences. For example, inconsistent compilation behaviors may hinder the migration of smart contracts across different blockchain platforms, potentially causing deployment failures. They may also mislead developers during contract development and debugging, since identical Solidity programs may produce different compilation results. In addition, such inconsistencies may introduce vulnerabilities that adversaries could exploit, potentially resulting in substantial financial losses. Therefore, systematic detection of these inconsistencies is critical for trustworthy smart contracts.

\para{Challenges} However, systematically detecting cross-compiler inconsistencies in Solidity is far from straightforward. 

First, generating programs that trigger inconsistencies is difficult. Existing fuzzers rely on coarse-grained mutation strategies that struggle to produce syntactically diverse programs. POLYGLOT~\cite{chen2021one} manipulates the program’s IR, Fuzzol~\cite{mitropoulos2023syntax} modifies AST nodes, and afl-compiler-fuzzer~\cite{groce2022making} disassembles and reassembles programs, often requiring numerous attempts to generate complex structures. Moreover, these mutations generally lack boundary-awareness, failing to target conditions that reveal subtle inconsistencies.

Second, identifying inconsistencies across compilers is nontrivial.  Existing tools either focus on detecting bugs in a single compiler \cite{chen2021one,groce2022making,mitropoulos2023syntax} or perform differential testing across multiple compilers that target the same execution environment and platform \cite{tian2024differential,tu2022detecting,ossfuzz}. However, Solidity compilers often target different execution environments, producing outputs that cannot be directly compared, which makes these existing approaches unsuitable for cross-compiler inconsistency detection.

\para{Our Insight} While Solidity compilers produce platform-specific bytecode, their interpretation of the language syntax should be consistent. We observe that compilers often diverge when handling corner cases involving numerous boundary conditions (see Section~\ref{sec:motivation} for details). These differing approaches are a primary source of \textit{cross-compiler inconsistencies}, the main targets for detection in this paper. Such inconsistencies manifest as differences where, for the same source code, one compiler might crash while another does not, one produces runnable bytecode while the other's output fails, or their resulting bytecodes execute with different outcomes.

\para{Our Approach} We propose \system{}, a novel framework for detecting cross-compiler inconsistencies in Solidity. The architecture of \system{} consists of three main components, namely a mutation rule generator, a fine-grained mutator, and an execution-based detector.

The fuzzing process begins with the mutation rule generator, which creates a rich set of mutation rules. The fine-grained mutator then adaptively applies these rules to seed programs to generate variants likely to expose compiler inconsistencies. Finally, the execution-based detector identifies inconsistencies by compiling and running these variants across different compilers, analyzing not only errors but also subtle differences in their execution outcomes.

Our evaluation shows that \system{} outperforms state-of-the-art fuzzers (i.e., POLYGLOT~\cite{chen2021one}, afl-compiler-fuzzer~\cite{groce2022making}, FUZZOL~\cite{mitropoulos2023syntax}, and solc\_ossfuzz~\cite{ossfuzz}) in both efficiency and efficacy. Specifically, \system{} achieves up to an 18$\times$ higher compilation success rate and 1.8$\times$ greater code coverage, demonstrating its ability to generate high-quality test programs. In terms of efficacy, \system{} detected 31$\times$ more inconsistencies than the compared fuzzers.
We applied \system{} to find inconsistencies among six popular Solidity compilers (i.e., Solc~\cite{solc}, Revive~\cite{revive}, Zksolc~\cite{zksolc}, Solang~\cite{solang}, Sold~\cite{sold}, and Solar~\cite{solar}). As a result, {\confirmedBugs} inconsistencies have been confirmed as bugs by the respective developers, and {\fixedBugs} of them have already been fixed.


In summary, this paper makes the following contributions:
\begin{itemize}
    \item  We present \system{}, the first fuzzing framework designed specifically to detect inconsistencies across Solidity compilers.
    \item We design a novel fuzzing strategies that combines a mutation rule generator with a fine-grained mutator to efficiently trigger cross-compiler inconsistencies.
    \item We implement an execution-based oracle, which identifies not only compilation and runtime errors but also subtle differences in execution outcomes.
    \item We demonstrate the effectiveness of \system{} through a comprehensive evaluation. \system{} outperforms state-of-the-art tools and has identified \bugs{} previously unknown inconsistencies across six popular Solidity compilers.
    
\end{itemize}

The rest of this paper is organized as follows. Section~\ref{sec:motivation} provides background information and motivating examples. Section~\ref{sec:overviewTool} and Section~\ref{sec:detailTool} describe the design and methodology of \system{}, followed by implementation details in Section~\ref{sec:implementation}. Section~\ref{sec:experiments} presents a comprehensive evaluation. Section~\ref{sec:discussion} discusses threats to validity and limitations. Section~\ref{sec:relateWork} reviews related work, and Section~\ref{sec:conclusion} concludes the paper. 
\section{Background and Motivation}\label{sec:motivation}
\begin{figure}[t]
\centering
\includegraphics[width=3.5in]{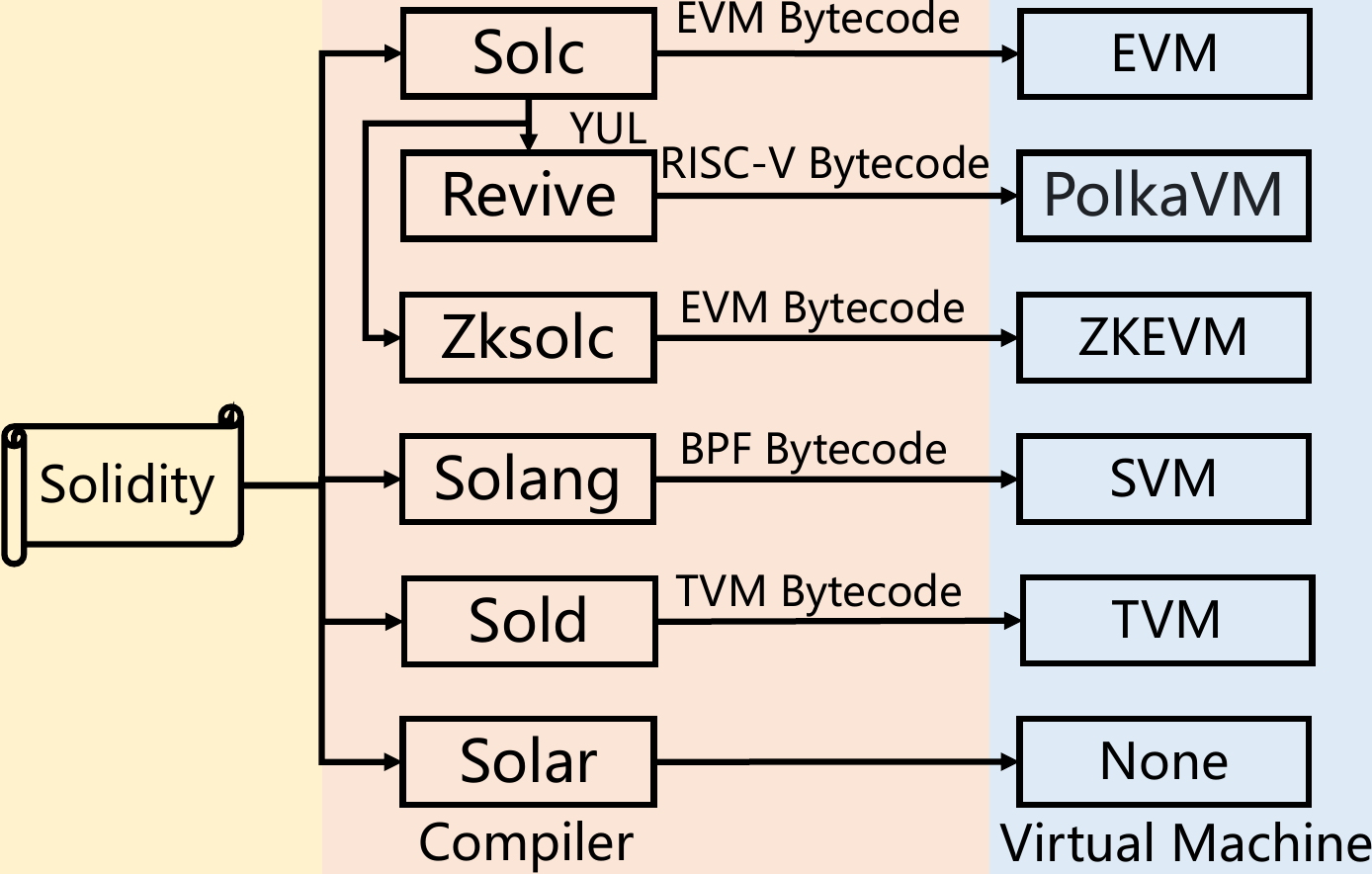}
\caption{The compilation and execution process under different Solidity compilers.}
\label{compilerFlow}
\vspace{-2ex}
\end{figure}

\subsection{Solidity Compilers}
Figure~\ref{compilerFlow} illustrates six widely used Solidity compilers. A Solidity compiler primarily translates source code into bytecode that can be executed on a virtual machine. In practice, this bytecode executes within a complex execution environment, which we refer to as the executor. In our work, the executor encompasses not only the VM itself, but also higher-level runtime modules such as transaction handling, contract calls, and state management. Our analysis evaluates the behavior of programs across this entire execution flow.

Inconsistencies across Solidity compilers can arise from two main sources. First, some compilers (e.g., Solc, Zksolc, Solang) target different blockchain platforms. To support each platform, they require customized designs that produce the bytecode needed for that environment. Second, even compilers targeting the same environment can adopt different internal designs. For instance, Solar implements aggressive optimizations to improve compilation efficiency, while its backend remains under active development. These factors can lead to observable differences in compilation and execution outputs, which \system{} detects as inconsistencies.

\subsection{Cross-Compiler Inconsistencies}

To systematically study these inconsistencies, we first analyzed existing compiler behaviors. Based on these observations, we propose a taxonomy that classifies cross-compiler inconsistencies into four types.

\begin{itemize}
    \item \textbf{Error Message Inconsistency (EMI).} Both compilers fail to compile the same code, but one provides a detailed error message (e.g., locations and causes) while the other does not. Vague error details from a compiler can significantly hinder debugging and degrade the developer experience.
    \item \textbf{Compilation Status Inconsistency (CSI).} One compiler successfully compiles the code, while another reports an error. This type of inconsistency can prevent a program that is valid in one blockchain ecosystem from being deployed in another.
    \item \textbf{Execution Status Inconsistency (ESI).} Bytecode from one compiler executes successfully, while bytecode from another fails at runtime. This can occur either when a compiler generates invalid bytecode for its target executor, or when the executor itself imposes inappropriate restrictions. Such failures often cause transactions to revert, impeding the migration of decentralized applications.
    \item \textbf{Execution Output Inconsistency (EOI).} Bytecode from both compilers executes without error but produces different results. This semantic inconsistency can subvert the program's intended logic and may lead to significant financial losses.

\end{itemize}

\subsection{Motivating Example}


Figure~\ref{structBugProgram} illustrates a case of compilation status inconsistency, where Solc accepts the program while Solang reports a compilation error. This compilation error originates in the \texttt{\small storage\_align} of Solang. As shown in Figure~\ref{structBugReason}, during the recursive calculation of storage alignment for structure fields, the function encounters a user-defined value type. Because the Solang compiler lacks a specific implementation for this case, it hits an \texttt{\small unimplemented!()} branch and fails. Overall, the compilation error is due to Solang's incomplete support for user-defined types within structures.

This motivating example shows that inconsistencies often arise from differing boundary conditions across compilers and executors, such as the unimplemented path in Figure~\ref{structBugReason}. Therefore, generating programs that trigger these boundary conditions can help uncover inconsistencies between compilers.

Beyond this, some inconsistencies do not directly trigger boundary conditions but still activate the surrounding code blocks. These inconsistencies arise from design differences among compilers and executors, and their exact causes are often difficult to trace in the source code. As a result, they typically require programs with diverse syntactic structures to be triggered. Since they can still reach code blocks containing boundary conditions, random mutation strategies targeting these code blocks can be employed to expose such inconsistencies.

Finally, we can conclude that triggering inconsistencies requires both the generation of syntactically diverse programs and the presence of boundary conditions in the source code.

\begin{figure}[htbp]
\begin{minipage}[t]{1\linewidth}
\begin{lstlisting}[language=Solidity,
    mathescape, 
    firstnumber=1, 
    escapechar=\%,
    linebackgroundcolor = 
    {\color{verylightgray}\ifnum\value{lstnumber}=3\color{githubpink}
    \else\fi
    }]
type MyValueType is uint;
contract X {
    struct S { MyValueType x; }
}
\end{lstlisting}
\centering
\caption{An example of compilation status inconsistency. Solc compiles the program successfully, whereas Solang reports a compilation error.}
\label{structBugProgram}
\end{minipage}
\begin{minipage}[t]{1\linewidth}
\begin{lstlisting}[language=Rust,
    mathescape, 
    firstnumber=1, 
    escapechar=\%,
    linebackgroundcolor =     {\color{verylightgray}\ifnum\value{lstnumber}=9\color{githubpink}
    \else\ifnum\value{lstnumber}=14\color{githubpink}
    \else\ifnum\value{lstnumber}=20\color{githubpink}
    \else\fi\fi\fi
    }
    ]
pub fn storage_align(&self, ns: &Namespace) -> BigInt {
    let length = match self {
        // ...
        Type::Array(ty, _) => {
            // ...
            ty.storage_align(ns)
            // ...
        }
        Type::Struct(s) => s
            .definition(ns)
            .fields
            .iter()
            .filter(|f| !f.infinite_size)
            .map(|f| f.ty.storage_align(ns))
            .max()
            .unwrap_or_else(|| 1.into()),
        Type::String | Type::DynamicBytes => // ...
        Type::InternalFunction { .. } =>     // ...
        // ...
        _ => unimplemented!(),
        // ...
    };
    // ...
}
\end{lstlisting}
\centering
\caption{The source code of the Solang compiler that handles struct}
\label{structBugReason}
\end{minipage}
\end{figure}

\subsection{Limitations of State-of-the-art Fuzzers}

Existing approaches \cite{chen2021one,groce2022making,mitropoulos2023syntax,ossfuzz} are primarily designed to detect bugs within individual Solidity compilers. When it comes to uncovering inconsistencies across different compilers, these approaches face two limitations. 

\para{Difficulty in Generating Inconsistency-Triggering Programs} First, the mutation strategies in existing fuzzers \cite{chen2021one,mitropoulos2023syntax,groce2022making} are relatively coarse-grained, making it difficult to generate syntactically diverse programs. Specifically, POLYGLOT\cite{chen2021one} inserts, deletes, and replaces elements in the program's intermediate representation (IR). Fuzzol\cite{mitropoulos2023syntax} modifies the nodes of the program's AST. afl-compiler-fuzzer\cite{groce2022making} disassembles programs and then reassembles them. These coarse mutation strategies rely on a large number of failed attempts before they can mutate relatively complex syntactic structures. 

Second, the mutation strategies lack boundary-orientation. As shown in Figure~\ref{structBugReason}, boundary conditions serve as an intuitive manifestation of certain inconsistencies. The mutation strategies of fuzzers \cite{chen2021one,groce2022making,mitropoulos2023syntax} are relatively generic, making it difficult to generate programs that trigger boundary conditions. Apart from this, LLM4CBI \cite{tu2024isolating} incorporates 13 mutation rules (e.g., inserting an if statement or a loop) for compiler bug isolation. DFUZZ \cite{zhang2025your} extracts edge cases from DL library APIs to guide program mutation. These fuzzers' mutation strategies focus on domain-specific problems rather than compiler inconsistencies, and therefore also lack boundary-orientation.

\para{Difficulty in Identifying Inconsistencies} Tools \cite{chen2021one,groce2022making,mitropoulos2023syntax} only support the detection of compilation error bugs. Therefore, these tools \cite{chen2021one,groce2022making,mitropoulos2023syntax} can only detect inconsistencies in the compilation results. DeSCDT \cite{tian2024differential} and yul\_proto\_diff\_ossfuzz\cite{ossfuzz} aim to detect optimization issues in the Solc compiler. They compare the execution results of programs compiled with and without optimizations to check for any differences. Therefore, these tools can only detect whether the program executes correctly in the EVM and cannot be extended to executors matching other compilers. 
\section{Overview of \system{}}\label{sec:overviewTool}

\begin{figure*}[t]
    \centering
    \includegraphics[width=4in]{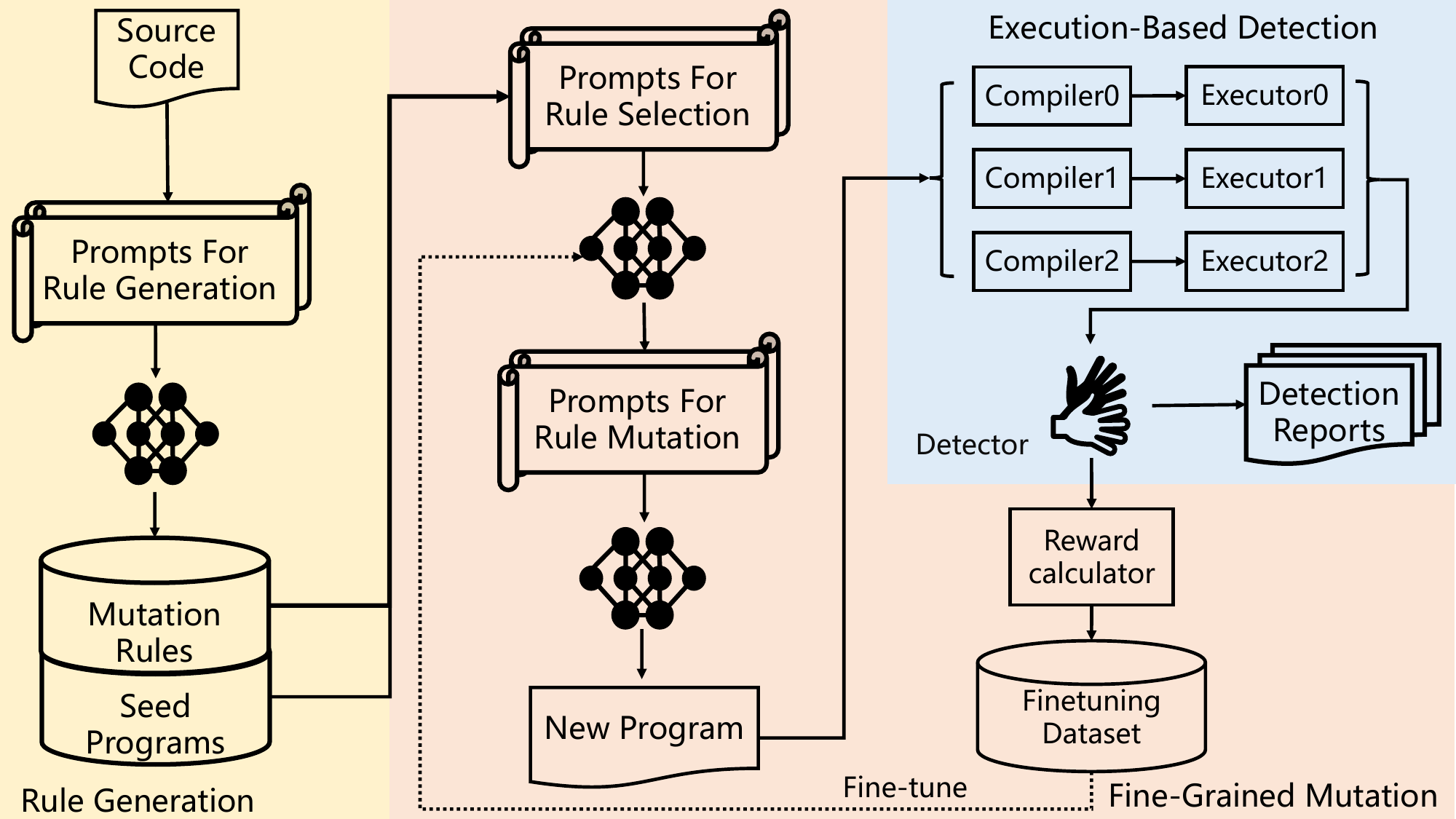}
    \vspace{-1ex}
    \caption{Workflow of \system{}.}
    \label{SoCCFuzzProcess}
\end{figure*}

\subsection{Our idea} This work presents \system{}, a fuzzer designed to automatically generate Solidity programs that reveal inconsistencies among Solidity compilers. Effective test programs exhibit two key properties, including syntactic diversity and a focus on boundary conditions.

Our approach combines two types of mutation rules to effectively generate test programs. Given boundary conditions in source code (i.e., compiler and executor), \system{} generates mutation rules based on syntactic features from the source code in order to achieve syntactic diversity. \system{} also generates mutation rules based on boundary conditions in order to trigger inconsistencies. The extraction is based on a Large Language Model (LLM) by analyzing source code.

To adaptively select mutation rules for fuzzing, \system{} first classifies all mutation rules and applies a recurring mutation loop to keep generating more promising test programs for triggering cross-compiler inconsistencies. This process in continuously optimized using reinforcement learning to improve its effectiveness over time.

Finally, \system{} detects inconsistencies by executing each generated program across multiple compilers and comparing the outcomes. To ensure an accurate comparison, a normalization step decodes and aligns the outputs from different executors, accounting for variations in their semantics.

\subsection{Workflow of \system{}} 

As illustrated in Figure~\ref{SoCCFuzzProcess}, the \system{} framework operates in three main stages: mutation rule generation, fine-grained mutation, and execution-based detection.

The mutation rule generation stage begins by identifying boundary conditions within the target compilers and executors. \system{} then analyzes the syntactic features that activate these conditions to automatically generate syntax-oriented (SO) mutation rules. These rules are subsequently refined into boundary-oriented (BO) rules designed to specifically trigger sensitive areas.

In the fine-grained mutation stage, \system{} adaptively mutates seed programs. It employs a mutation-based fuzzing loop to keep generating more promising test programs from a given seed program. This process is continuously optimized via reinforcement learning, which uses feedback from the compilation and execution of new variants to improve future choices.

The final execution-based detection stage identifies inconsistencies through differential analysis. Newly generated programs are compiled and run across multiple toolchains. We apply semantic alignment and compare testing output. After filtering out false positives, programs that trigger inconsistencies are documented and reported.

\subsection{Fuzzing Scope}\label{Scope}

We identify the following six popular Solidity compilers as the subjects of our study: Solc, Revive, Zksolc, Solang, Sold, and Solar. According to DefiLlama~\cite{DefiLlama}, the Total Value Locked (TVL) of these compilers reaches 100 billion dollars as of August 2025. These compilers represent blockchain platforms with one of the highest Total Value Locked (TVL) records, making the security of these compilers significant. Among them, although the Solar compiler does not yet have a backend, it is planned to support EVM in the future \cite{introducingSolar}.

Building upon this selection, our research focuses on identifying four types of inconsistencies (i.e., EMI, CSI, ESI, EOI) across the six Solidity compilers. Table~\ref{inconsistencySupport} summarizes the types of inconsistency detection supported by \system{} for each compiler. Notably, the program execution environment is missing for Sold, and Solar currently does not support bytecode generation. Thus, neither compiler can detect inconsistencies related to execution results.

\begin{table}[t]
  \centering
  \caption{Inconsistency detection supported by \system{}.}
  \resizebox{0.48\textwidth}{!}{ 
    \begin{tabular}{lllllll}
      \toprule
      \multirow{2}[4]{*}{Inconsistency} & \multicolumn{6}{c}{Compiler} \\
      \cmidrule{2-7}          
      & Solc  & Revive & Zksolc & Solang & Sold  & Solar \\
      \midrule
      \multicolumn{1}{p{8.625em}}{Error Message} & \checkmark & \checkmark & \checkmark & \checkmark & \checkmark & \checkmark \\
      Compilation Status & \checkmark & \checkmark & \checkmark & \checkmark & \checkmark & \checkmark \\
      Execution Status & \checkmark & \checkmark & \checkmark & \checkmark &  - & - \\
      Execution Output & \checkmark & \checkmark & \checkmark & \checkmark &  - & - \\
      \bottomrule
    \end{tabular}
  }
  \label{inconsistencySupport}
\end{table}

\section{Details of \system{}}\label{sec:detailTool}

\subsection{Mutation Rule Generation}\label{subsec:ruleGeneration}
Two sets of mutation rules are generated: syntax-oriented rules and boundary-oriented rules. As for boundary-oriented rules, \system{} extracts boundary conditions from the source code of the compiler and executor and uses boundary conditions to generate rules. 
 
There are two main challenges in generating boundary-oriented mutation rules. (1) mutation rules need to be clear. The most straightforward approach is to let the LLM directly analyze the source code containing boundary conditions and generate mutation rules. However, such naively generated mutation rules are often vague, such as \texttt{"inserting unsupported types"}. Due to the wide variety of Solidity types, the LLM struggles to determine the specific type for mutation rules. DFUZZ\cite{zhang2025your} addresses this by listing all possible types, allowing the LLM to choose the appropriate one for edge cases. However, compiler boundary conditions are more complex and lack a fixed format, making it difficult for the LLM to select the correct type. Additionally, mutation rule ambiguity can arise from vague descriptions, like \texttt{"increase the complexity of nested types"}. DFUZZ can only complete specific types but struggles to handle vague concepts like \texttt{"complexity"}. (2) mutation rules need to be comprehensive. A piece of source code may contain multiple boundary conditions, and each boundary condition can have various triggering methods. Mutation rules generated by LLMs through direct analysis of the code often miss many ways to trigger these boundary conditions. 

\begin{figure}[t]
\centering
\includegraphics[width=3.3in]{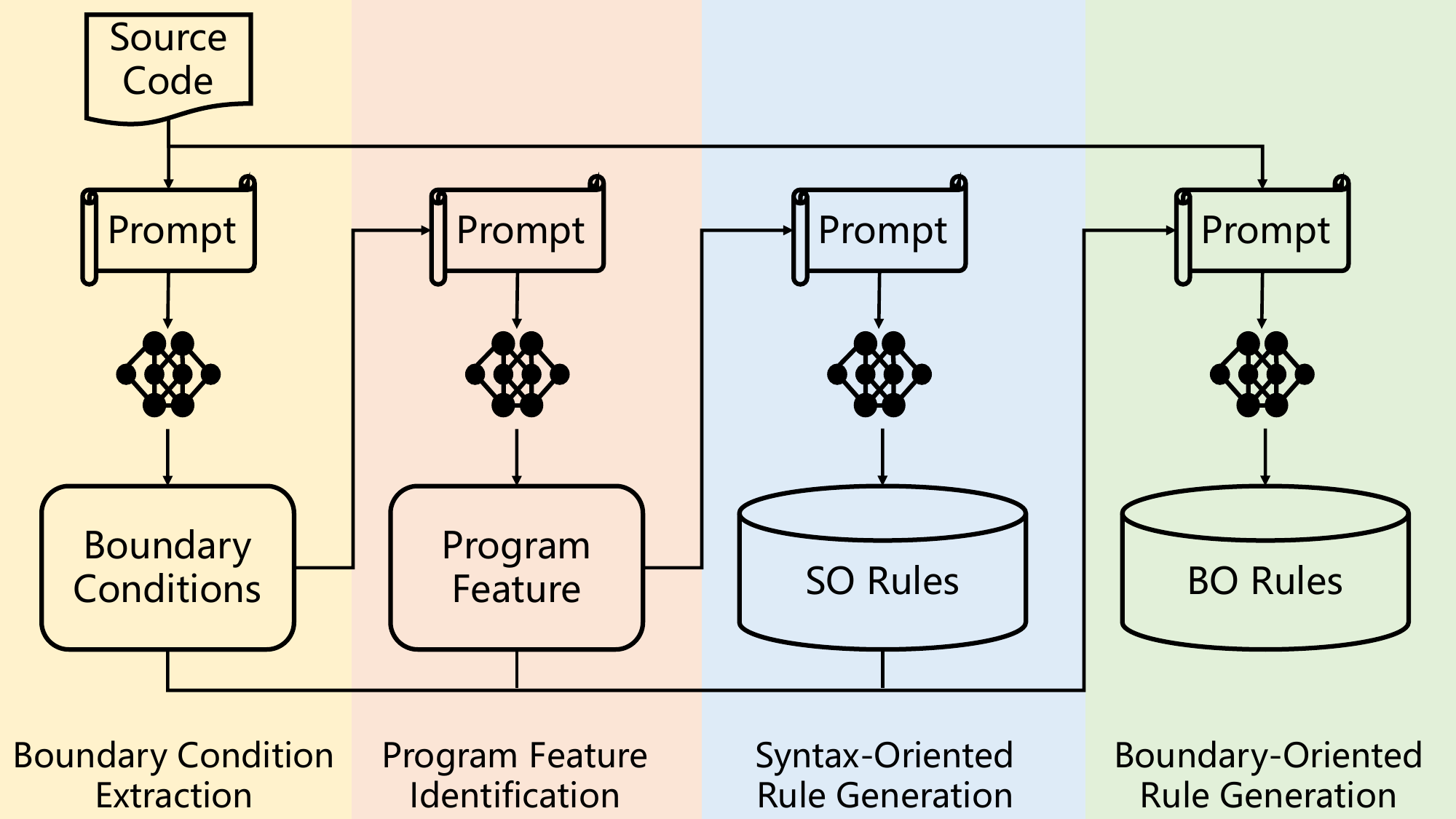}
\vspace{-1ex}
\caption{The process of rule generation.}
\label{ruleGenerationProcess}
\end{figure}

To address the above challenges, \system{} adopts a stepwise approach to enhancing mutation rules, shown in Figure~\ref{ruleGenerationProcess}. First, once the code blocks containing boundary conditions are obtained, \system{} identifies the program features that can trigger these code blocks and generate simple mutation rules for each feature (syntax-oriented rules). Building on these features and syntax-oriented rules, \system{} then creates more complex mutation rules (boundary-oriented rules).

To ensure these rules are effectively applied, \system{} handles two key issues. Specifically, to address the problem of vague rules in Challenge (1), \system{} embeds syntax-oriented rules into the prompt for boundary-oriented rule generation, enabling the LLM to either select one as a new rule or generate a new one based on it. Since syntax-oriented rules are inherently simple and clear, the generated rules also retain clarity. To address the problem of missing rules in Challenge (2), \system{} embeds one boundary condition and one feature into the prompt at a time, ensuring that the generated rule targets the feature and triggers the corresponding boundary condition.

\para{Boundary Condition Extraction} This step aims to extract boundary conditions present in the source code.

Boundary conditions can be explicit or implicit. Explicit boundary conditions include Exception and Error Handling, Unimplemented Features, Code Path Validity Checks, and Assertions. These usually have fixed identifiers such as \texttt{"error"}, \texttt{"panic"}, \texttt{"unimplemented"}, \texttt{"unreachable"}, or \texttt{"assert"}. For example, Figure~\ref{structBugReason} illustrates a code block containing the identifier \texttt{"unimplemented"}. Therefore, we can manually extract code blocks containing these identifiers. Implicit boundary conditions, such as type casting and memory out-of-bounds, are harder to identify manually. For these, we use LLM to assist extraction. Finally, we obtain code blocks containing boundary conditions.

\para{Program Feature Identification} This step aims to summarize the language features that can trigger code blocks containing boundary conditions. As shown in Figure~\ref{structBugReason}, the compiler source code reveals that boundary conditions are typically triggered through complex control flows. This reliance makes it difficult for the LLM to directly generate programs that can reach such conditions. Therefore, we first summarize the program features associated with code blocks containing boundary conditions. Building on these features, the LLM can generate precise mutation rules to trigger boundary conditions.

We directly employ the LLM to analyze the code blocks containing boundary conditions and identify the involved Solidity language features. For example, in Figure~\ref{structBugReason}, the LLM identifies features such as \texttt{fixed-size array}, \texttt{struct}, and \texttt{internal function}, among others. 

\para{Syntax-Oriented Rule Generation} This step aims to generate syntax-oriented mutation rules, which are used to randomly mutate programs containing the aforementioned features. Since the mutation rules obtained by directly analyzing boundary conditions with the LLM can be vague, syntax-oriented rules also serve as references to guide the creation of more precise boundary-oriented rules.

\begin{figure}[t]
\centering
\includegraphics[width=3.3in]{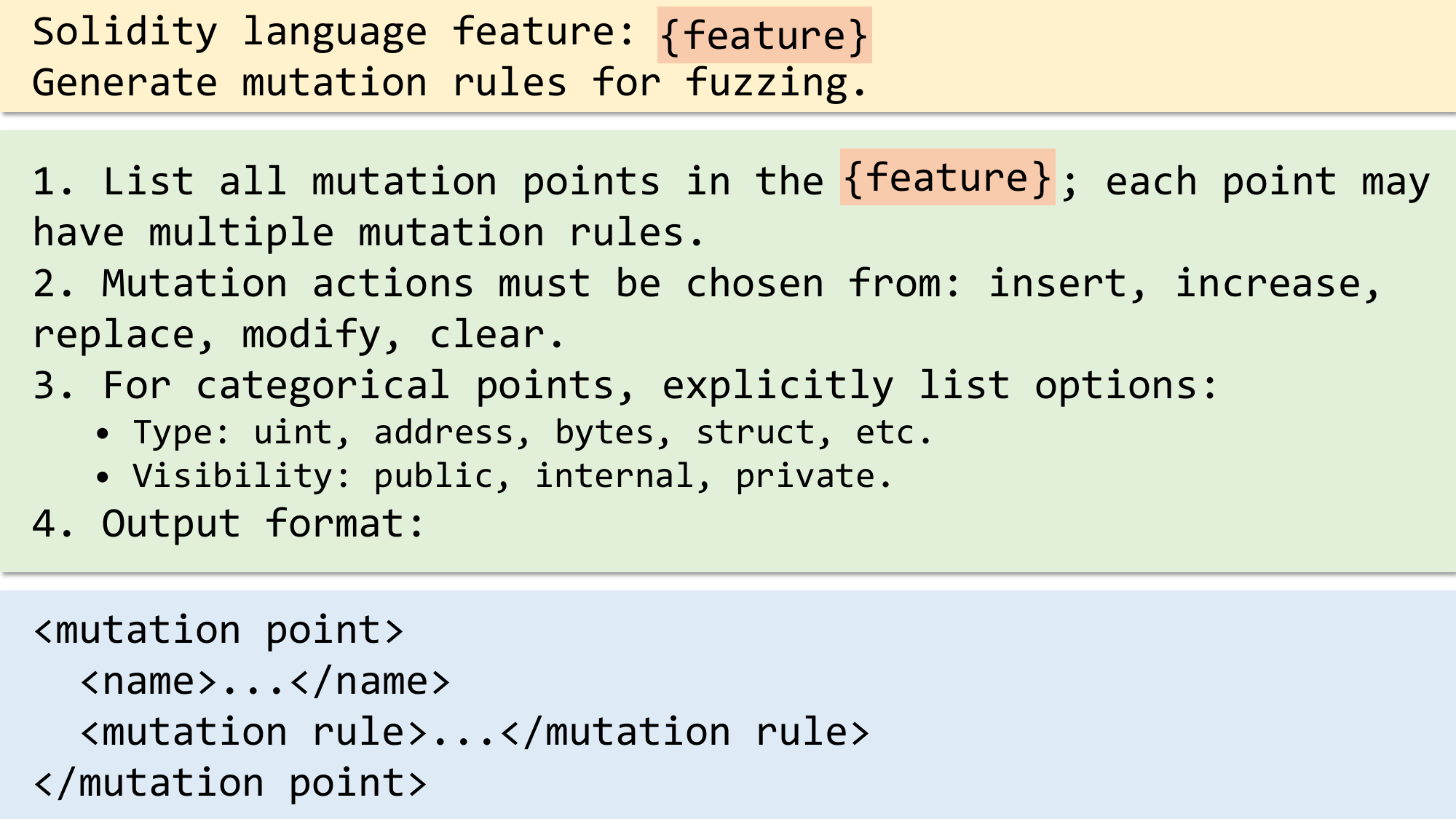}
\vspace{-1ex}
\caption{Prompt for generating syntax-oriented mutation rules}
\label{basicRulePrompt}
\end{figure}

Figure~\ref{basicRulePrompt} illustrates the prompt used for generating syntax-oriented rules. As each language feature consists of several mutation points, the first command in the prompt directs the LLM to list the mutation points associated with that feature. For instance, the struct includes mutation points such as its field types, visibility, and declaration location. As for mutation actions, we choose five representative types (insert, increase, replace, modify, and clear) as options for the LLM to select. 

\para{Boundary-Oriented Rule Generation}  
This step aims to generate boundary-oriented mutation rules based on program features and syntax-oriented rules. A single code block may contain multiple boundary conditions, and each boundary condition can be triggered in multiple ways. Each triggering method can correspond to a mutation rule. To ensure the completeness of mutation rules, we select only one feature and one boundary condition in each prompt, thereby addressing Challenge 2. At the same time, to ensure the clarity of mutation rules, we use syntax-oriented rules to guide the generation of boundary-oriented rules, thereby addressing Challenge 1.

\begin{figure}[t]
\centering
\includegraphics[width=3.3in]{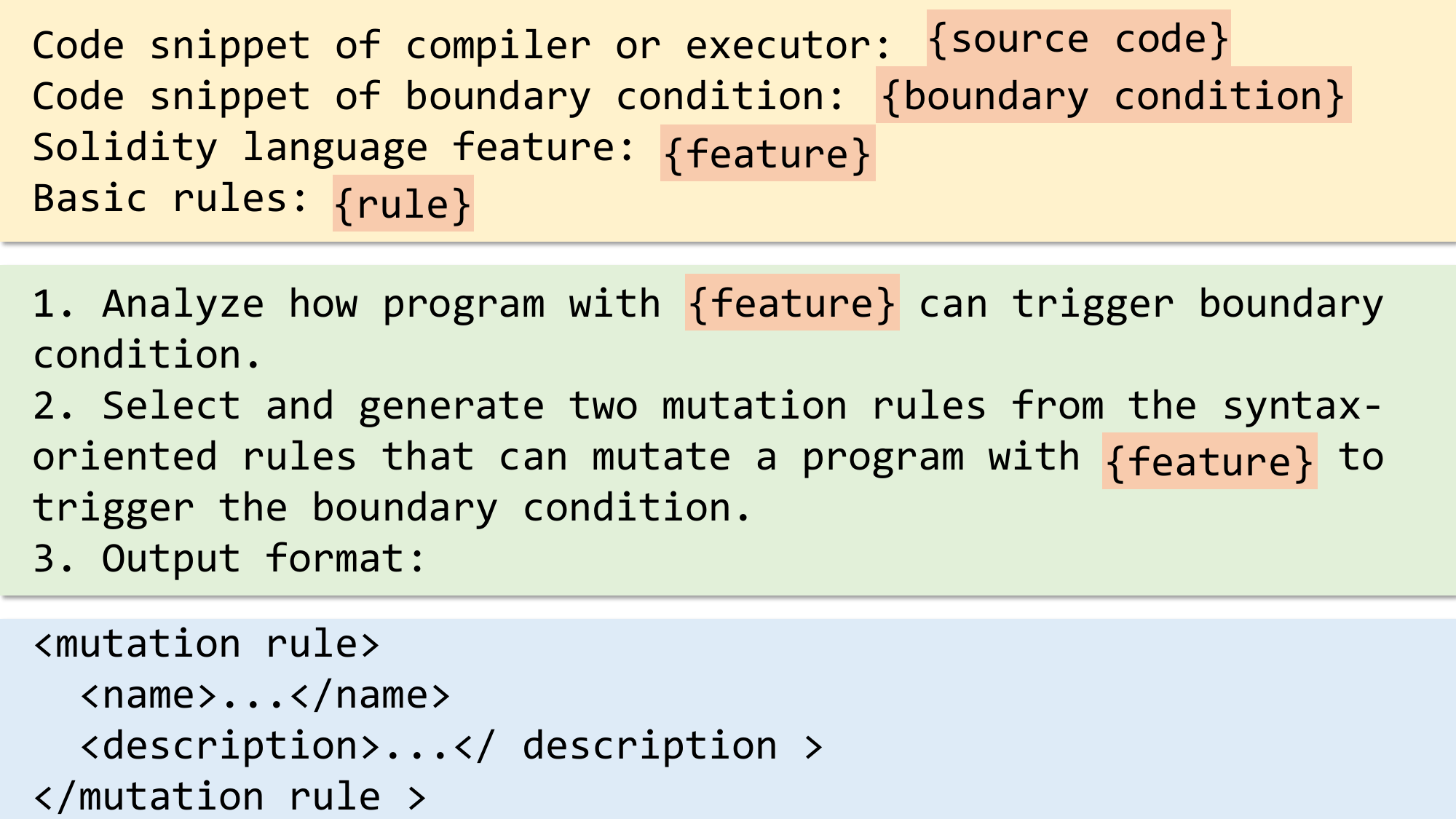}
\vspace{-1ex}
\caption{Prompt for generating boundary-oriented mutation rule}
\label{boundaryRulePrompt}
\end{figure}

Figure~\ref{boundaryRulePrompt} illustrates the prompt used for generating the boundary-oriented mutation rule. In the prompt, the first task asks the LLM to analyze how the current boundary-related feature can trigger the specified boundary condition. The second task asks the LLM to either select from existing syntax-oriented rules or synthesize new ones based on them.

After obtaining the rules, we need to categorize them. We use the features embedded in the prompt as labels, representing the major category to which each generated rule belongs. These features will be used in the subsection~\ref{subsec:programMutation} for rule selection.

\subsection{Fine-Grained Mutation}\label{subsec:programMutation}
After generating the mutation rules, we select the appropriate ones to apply mutations to the program. The process can be divided into three stages, namely rule selection, rule application, and selection optimization. During rule selection, \system{} selects the most suitable mutation rules for each seed program. During rule application, \system{} applies the selected rules to mutate the program and then fixes programs that fail to compile. During selection optimization, \system{} leverages reinforcement learning to fine-tune the LLM, optimizing the rule selection process.

\para{Rule Selection} \system{} employs two sets of mutation rules to perform mutations. Boundary-oriented rules are designed to uncover inconsistencies related to boundary conditions, while syntax-oriented rules ensure syntactic diversity of programs, thereby triggering inconsistencies related to the code blocks where boundary conditions reside. 

\begin{figure}[t]
\centering
\begin{subfigure}[t]{3.3in}
    \centering
    \includegraphics[width=\linewidth]{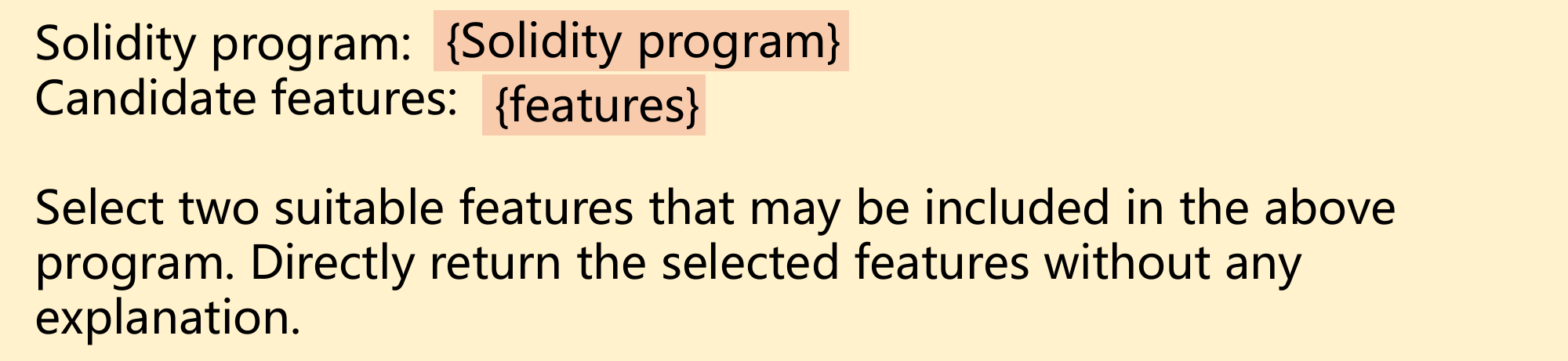}
    \caption{Prompt for selecting features.}
    \label{ruleSelectionPrompt0}
\end{subfigure}
\begin{subfigure}[t]{3.3in}
    \centering
    \includegraphics[width=\linewidth]{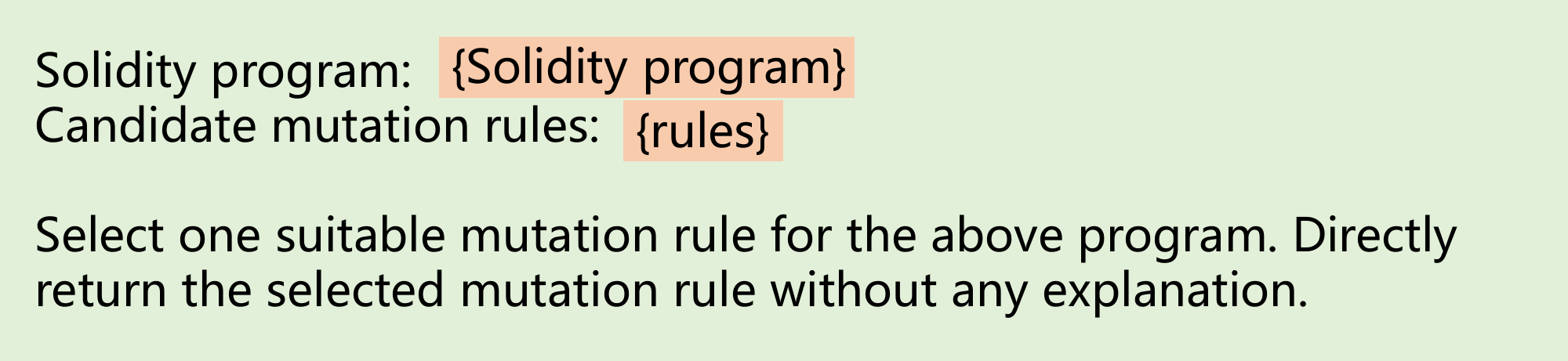}
    \caption{Prompt for selecting mutation rule.}
    \label{ruleSelectionPrompt1}
\end{subfigure}
\caption{Prompt for selecting features and mutation rule.}
\label{ruleSelectionPrompt}
\vspace{-3ex}
\end{figure}

\begin{figure}[t]
\centering
\begin{subfigure}[t]{3.3in}
    \centering
    \includegraphics[width=\linewidth]{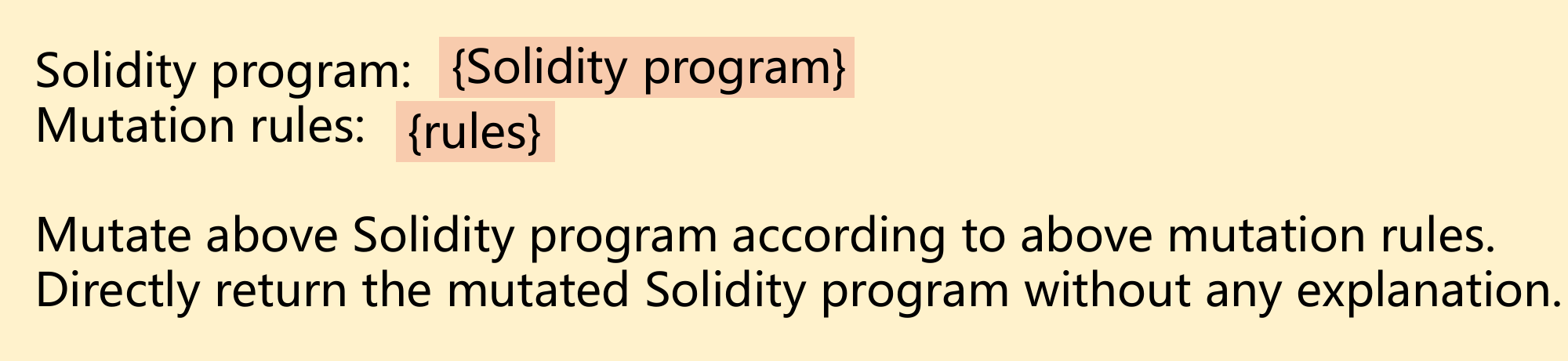}
    \caption{Prompt for mutating program.}
    \label{programMutationPrompt0}
\end{subfigure}
\begin{subfigure}[t]{3.3in}
    \centering
    \includegraphics[width=\linewidth]{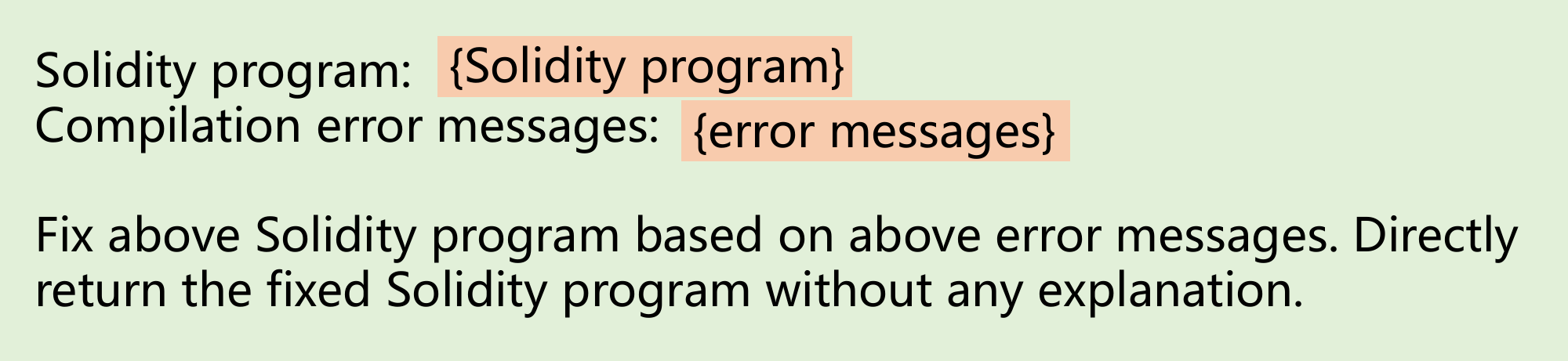}
    \caption{Prompt for repairing program.}
    \label{programMutationPrompt1}
\end{subfigure}
\caption{Prompt for mutating and repairing program.}
\label{programMutationPrompt}
\vspace{-3ex}
\end{figure}

Figure~\ref{programMutationPrompt} shows the prompt for selecting features and mutation rules. \system{} first prompts the LLM to identify which features the seed program contains (Figure~\ref{ruleSelectionPrompt0}), and then selects a specific mutation rule under the chosen feature (Figure~\ref{ruleSelectionPrompt1}). Since mutation rules are based on the program already having certain features, this ensures that the selected rules are well-suited to the seed program.

\para{Rule Application} Figure~\ref{programMutationPrompt} shows the prompt for mutating and repairing programs. After selecting the appropriate rules, \system{} applies them to mutate the seed program (shown in Figure~\ref{programMutationPrompt0}). If the generated program fails to compile with Solc, \system{} provides the error message to the LLM to repair the program (shown in Figure~\ref{programMutationPrompt1}).

While the repair process may alter the program's original semantics, it helps uncover more cross-compiler inconsistencies. Programs generally need to be compilable with Solc in order to potentially trigger inconsistencies. Programs that cannot compile with Solc often also fail on other Solidity compilers and therefore cannot reveal cross-compiler differences. Repairs also enable the generation of programs with complex syntactic structures that exercise boundary conditions and are more likely to reveal inconsistencies. Generating such complex structures in a single mutation step can be challenging for the LLM, so iterative repair is applied to ensure successful program mutations.

\para{Rule Selection Optimization} To optimize the rule selection process, \system{} leverages reinforcement learning to fine-tune the LLM. Inspired by previous studies \cite{eom2024fuzzing,tu2024isolating}, we finetune our model using the GRPO algorithm \cite{shao2024deepseekmath}. Compared to the PPO \cite{schulman2017proximal} algorithm used in CovRL-Fuzz\cite{eom2024fuzzing}, GRPO does not require a separate value model, saving both memory and computational cost. Additionally, \system{} does not adopt a reward model like the one in CovRL-Fuzz\cite{eom2024fuzzing}; instead, it uses a reward function to score the outputs of the LLM. Since the reward function itself can produce precise scores, the step of training a reward model using the reward function can be eliminated. 

The reward function uses the results of inconsistency detection to calculate the score of the mutated program. This score indirectly reflects the quality of the selected mutation rules. The score is calculated as follows.

\begin{equation}
\text{score} =
\begin{cases}
1, & \text{if an inconsistency occurs.} \\
-1, & \text{if the inconsistency is a false positive.}  \\
-1, & \text{if Solc fails to compile} \\
R_{\text{covDiff}}, & \text{otherwise.}
\end{cases}
\end{equation}

If the new program triggers an inconsistency, the score is 1. If the inconsistency is a false positive, a penalty is applied with a score of -1. If no inconsistency is triggered and Solc fails to compile the program, a penalty of -1 is also applied. Otherwise, the new program is successfully compiled by Solc but does not trigger any inconsistencies, and the score is determined by the change in coverage. The coverage difference is calculated as follows: compile the programs before and after mutation using Solc separately, measure the coverage within the Solc compiler for each version, and then compute the difference. The coverage difference ranges from -1 to 1, which helps guide the LLM to generate programs with richer syntax through mutation.

\subsection{Execution-Based Detection}\label{subsec:inconsistencyDetection}
To uncover inconsistencies, we compare the compilation and execution results of programs across different compilers. While using multiple compilers introduces intentional design differences, \system{} ensures the reliability of its differential oracle through two mechanisms. First, we select Solc as the baseline compiler, as it is the most widely used reference compiler for Solidity. Second, we systematically eliminate inconsistencies caused by documented design differences using compiler-provided documentation and pattern matching, so that only unexpected or potentially erroneous behaviors are detected. This combination allows \system{} to distinguish true inconsistencies from expected behavior in a reproducible manner.

\para{Compilation Result Comparison} As shown in subsection \ref{Scope}, compilation result inconsistencies include error message inconsistency and compilation status inconsistency. Both inconsistencies can be detected by directly comparing the outputs of the compilers.

Error message inconsistency means that both compilers produce errors, but one outputs a clear error message while the other does not. A clear error message typically begins with the keyword "error," followed by the location in the program that caused the error and the reason for the failure. Therefore, \system{} can identify clear error messages by matching keywords.

Compilation status inconsistency means that one compiler can compile the code successfully while another fails to compile it. This inconsistency can be detected by directly comparing the outputs of the compiler. When compilation succeeds, the compiler produces bytecode; when it fails, the compiler throws an error.

\para{Execution Result Comparison} Execution result inconsistencies include execution status inconsistency and execution output inconsistency. Both inconsistencies can be detected by directly comparing the outputs of the executors. 

Execution status inconsistency means that a program can be successfully executed after being compiled by one compiler but fails after being compiled by another. If the executor produces an error, it indicates a runtime failure; otherwise, the program executes successfully.

Execution output inconsistency means that the outputs of the executor differ after executing the two programs. As shown in Table~\ref{inconsistencySupport}, \system{} supports detecting execution result inconsistencies for four compilers (i.e., Solc, Revive, Zksolc, Solang). Among them, only the virtual machine (i.e., SVM) paired with Solang has a different output format compared to the other virtual machines (i.e., EVM, PolkaVM, ZKEVM). The output of SVM is in borsh \cite{borsh} format, while the outputs of the other three virtual machines are in ABI-encoded \cite{abi} format. For the former, \system{} extracts specific data from the Borsh class. For the latter, \system{} uses ABI decoder to decode the output. The result is two outputs that can be compared.

\para{Comparison Target Selection} Due to varying levels of compatibility between different Solidity compilers and Solc, customized handling is required when performing comparisons. 

As shown in Figure~\ref{compilerFlow}, both Revive and Zksolc rely on Solc to generate Yul programs, which are then compiled downstream. Solar aims to optimize the performance of Solc. Therefore, these three compilers (i.e., Revive, Zksolc, and Solar) are highly compatible with Solc and are expected to produce the same compilation and execution results, making direct comparison feasible. 

Solang and Sold have limited compatibility with Solc and do not support many Solidity language features. As a result, a large number of programs can be successfully compiled and executed with Solc, whereas Solang and Sold tend to trigger errors. Therefore, to eliminate such meaningless inconsistencies, we apply the following restriction. As for compilation status inconsistency, only when the error messages from Solang and Sold are unclear do we consider the inconsistency to be valid. As for execution status inconsistency, we only consider cases where the program executes correctly when compiled with Solang, but fails during execution when compiled with Solc.




\para{False Positive Elimination} 
This part aims to eliminate false positives among the detected inconsistencies. Each Solidity compiler provides a document describing its design-level differences from Solc (e.g., Revive~\cite{reviveDiff}, Zksolc~\cite{zksolcDiff}, Solang~\cite{solangDiff}, Sold~\cite{soldDiff}), specifically indicating which syntactic features are handled differently and the expected compilation or execution outcomes.

We consider an inconsistency a false positive if the responsible syntactic feature has a clearly defined expected outcome in the compiler's document. Features mentioned without precise expected behavior are not filtered, ensuring that potential new or subtle inconsistencies remain detectable.

To implement this, we manually extract syntactic features from these documents and encode them as patterns. During inconsistency detection, \system{} performs automated pattern matching on input programs: if a program matches a documented feature pattern, the corresponding inconsistency is filtered out. This semi-automated approach reduces false positives while preserving novel or unexpected behaviors, providing a reproducible and reliable method.

\section{Implementation}\label{sec:implementation}
\para{Implementation of \system{}} We use Python to implement \system{}. The compiler versions used in \system{} are as follows: Solc\cite{solc} (0.8.28), Revive\cite{revive} (v0.1.0-dev.9), Zksolc\cite{zksolc} (v1.5.10), Solang\cite{solang} (v0.3.3), Sold\cite{sold} (0.77.0), and Solar\cite{solar} (0.1.1). The execution environments for Revive and Solang are derived from modifications to the test environments (\cite{reviveRunner,solangRunner}) within their respective compiler projects. These test environments are originally designed for unit tests where Solidity contracts are defined directly in the test code. To support differential testing in \system{}, we adapt them to execute arbitrary Solidity programs. Specifically, we extend the environments to accept the Solidity file, contract name, and function name as inputs. We also normalize runtime outputs to enable automated comparison across compilers. For the other compilers, Zksolc and Sold directly use their respective node-based execution environments\cite{era-test-node,ever-node}. However, due to a bug in Sold's node, it cannot run correctly. Therefore, in subsequent experiments, we do not test execution-related inconsistencies for Sold. In addition, Solar does not support bytecode generation. Therefore, execution-related inconsistencies for Solar are not considered. 

As for LLM, we use ChatGPT-4o to generate mutation rules, Qwen2.5-Coder-7B-Instruct\cite{Qwen2.5-Coder-7B-Instruct} to select features, Qwen2.5-Coder-0.5B-Instruct\cite{Qwen2.5-Coder-0.5B-Instruct} to select mutation rules, and Qwen-coder-plus\cite{qwen-coder-plus} to apply the mutations to the program. 

The reasons for selecting the LLM are as follows. rule generation requires a model capable of analyzing source code, such as ChatGPT-4o. Feature selection demands the LLM to identify key features of a program; tests show that Qwen models under 7B parameters cannot meet this need. Rule selection is size-independent, so a smaller model is used for easier fine-tuning. Program mutation, requiring strong code understanding, is suited for models like Qwen-coder-plus.

\para{Implementation of compared fuzzers} We selected the following fuzzers for comparison: POLYGLOT \cite{chen2021one}, afl-compiler-fuzzer \cite{groce2022making}, FUZZOL \cite{mitropoulos2023syntax}, solc\_ossfuzz of OSS-FUZZ \cite{ossfuzz}. We first introduce the characteristics of these fuzzers. As for program mutation, POLYGLOT modifies the IR of the program, Fuzzol alters the AST of the program, solc\_ossfuzz applies string-level mutations (bit flipping, byte insertion, and deletion), and afl-compiler-fuzzer decomposes and reassembles the program. For the bug detection module, POLYGLOT, afl-compiler-fuzzer, and FUZZOL all use compilation crashes as oracles. solc\_ossfuzz determines whether the program compiled by Solc can execute correctly. Therefore, in our task, these fuzzers can only detect inconsistencies in the compilation results.

The configuration for these fuzzers is as follows: The timeout is set to two seconds for all fuzzers. afl-compiler-fuzzer includes two program mutation methods (text-mutation and splice-mutation). We set their triggering probabilities to 50\% each. 

\section{Evaluation}\label{sec:experiments}
This section presents a comprehensive evaluation of \system{}, guided by the following research questions:
\begin{itemize}
    \item \textbf{RQ1:} How effective is \system{} at detecting inconsistencies across different Solidity compilers?
    \item \textbf{RQ2:} How does the performance of \system{} compare to state-of-the-art Solidity fuzzers?
    \item \textbf{RQ3:} What is the contribution of each of the main components of \system{} to its overall effectiveness?
\end{itemize}

\para{Environment Setup} 
All experiments were conducted on a Linux server equipped with an Intel Xeon Platinum 8360H CPU (3.00GHz) and 125 GB of RAM, running Ubuntu 22.04. An NVIDIA GeForce RTX 4090 GPU was used for local model inference and fine-tuning of Qwen2.5-Coder-0.5B-Instruct. All other LLMs were accessed via their respective APIs.

\para{Dataset} 
Our initial benchmark dataset was constructed from three sources: the dataset from Ma et al.~\cite{ma2024towards}, the official Solidity historical bugs~\cite{sol_his_bug}, and the official Solidity test suite~\cite{solc_test}. This resulted in a corpus of 4,260 Solidity programs that successfully compile with the Solc compiler.

\subsection{Inconsistencies Identification} 
As shown in Table~\ref{inconsistencySimpleResult}, \system{} has successfully identified \bugs{} inconsistencies across six Solidity compilers, including 15 from Revive, 9 from Zksolc, 34 from Solang, 4 from Sold, and 2 from Solar. Detailed information on the inconsistencies can be found in our repository~\cite{ParityFuzz}. All identified inconsistencies have been reported, with approximately one-third confirmed by the developers. At the time of writing, \fixedBugs{} inconsistencies had been fixed by developers, and inconsistencies related to Revive received a bug bounty from the developers.

Meanwhile, we have made the following observations. Firstly, Solang has the most inconsistencies of all types. In Table~\ref{inconsistencySimpleResult}, its ESI count is only 3 because we only counted cases where Solc failed but Solang did not. The reverse cases were too many to include. The high inconsistency rate is mainly due to limited maintenance. Among the many inconsistencies we reported, only two issue \cite{solang_decode,solang_delete} has been acknowledged and fixed. Secondly, inconsistent execution results can be resolved not only by fixing the executor, but also by applying changes to the compiler or improving the documentation. For instance, due to ZKEVM not supporting \texttt{ripemd160}, the execution of \texttt{ripemd160("")} returns 0. To fix this issue \cite{zksolc_ripemd160}, the developers chose to throw an error during compilation. Another example is that Solc defaults to the EVM legacy codegen, whereas Zksolc uses the Yul codegen by default. This discrepancy may cause different execution results for certain programs. To resolve the issue \cite{zksolc_codegen}, the developers clarified this behavior in the documentation.

In the following case studies, we analyze representative inconsistencies to demonstrate how \system{} detects them and to examine their associated security consequences.

\begin{table}[t]
  \caption{Inconsistency detection results of \system{}.}
  \centering
  \resizebox{0.48\textwidth}{!}{  
    \begin{tabular}{cccccccc}
      \toprule
      \multirow{2}[4]{*}{Compiler} & \multicolumn{4}{c}{Inconsistency} & \multicolumn{3}{c}{Total} \\
      \cmidrule(lr){2-5}\cmidrule(lr){6-8}
            & EMI   & CSI   & ESI   & EOI   & Reported & Confirmed & Fixed \\
      \midrule
      Revive & 0     & 2     & 7     & 6     & 15    & 15    & 5 \\
      Zksolc & 0     & 0     & 4     & 5     & 9     & 8     & 2 \\
      Solang & 0     & 23    & 3     & 8     & 34    & 2     & 2 \\
      Sold   & 0     & 4     & 0     & 0     & 4     & 0     & 0 \\
      Solar  & 0     & 2     & 0     & 0     & 2     & 2     & 2 \\
      Total  & 0     & 31    & 14    & 19    & 64    & 27    & 11 \\
      \bottomrule
    \end{tabular}
  }
  \label{inconsistencySimpleResult}
\end{table}

\para{Case Study 1: Delegating a Call to an Invalid Address in Revive} As shown in Figure~\ref{delegatecallBugProgram}, the program fails to execute when compiled with Revive, but runs successfully when compiled with Solc. Figure~\ref{delegatecallBugReason} illustrates how the executor of Revive handles delegatecall. As we can see, for invalid addresses, the executor of Revive throws an error, causing the program execution to fail.

The way \system{} detects this inconsistency is as follows. In rule generation \ref{subsec:ruleGeneration}, \system{} uses LLM to analyze the source code in Figure~\ref{delegatecallBugReason} and generates the mutation rule: \texttt{"replace delegatecall with invalid contract type"}. Then, during the rule selection phase, \system{} successfully selects this rule for mutation. This mutation rule transforms the delegatecall target from a variable passed as a parameter (Figure~\ref{delegatecallOriginalProgram}, line 3) into an invalid address (Figure~\ref{delegatecallBugProgram}, line 4), thereby triggering the inconsistency. Therefore, \system{} can extract the boundary conditions in the source code very well and generate corresponding mutated rules to implement mutation.

This inconsistency can cause critical functionality to be blocked in real-world deployments. A contract that behaves correctly on Ethereum (compiled with Solc) may exhibit ambiguous behavior when migrated to Polkadot (compiled by Revive). If there are critical operations following a delegatecall, Revive may fail the delegatecall due to an invalid address, preventing any subsequent operations from being executed. This behavior can potentially be exploited by an attacker to block contract functionality or launch a DoS attack.

\begin{figure}[t]
\begin{minipage}[t]{1\linewidth}
\begin{lstlisting}[language=Solidity,
    mathescape, 
    firstnumber=1, 
    escapechar=\%,
    linebackgroundcolor = 
{\color{verylightgray}\ifnum\value{lstnumber}=3\color{githubgreen}
    \else\fi
    }]
contract C {
 function delegateToLibrary(address libraryAddress) external returns (bool) {
  (bool success, ) = libraryAddress.delegatecall(
    abi.encodeWithSignature("targetFunction()"));
  return success;
 }
}
\end{lstlisting}
\centering
\caption{The original version of the program in Figure~\ref{delegatecallBugProgram}}
\label{delegatecallOriginalProgram}
\end{minipage}
\begin{minipage}[t]{1\linewidth}
\begin{lstlisting}[language=Solidity,
    mathescape, 
    firstnumber=1, 
    escapechar=\%,
    linebackgroundcolor = 
    {\color{verylightgray}\ifnum\value{lstnumber}=4\color{githubpink}
    \else\fi
    }]
contract C {
 function delegateToLibrary(address libraryAddress) external returns (bool) {
  address invalidAddress = address(0x1);
  (bool success, ) = invalidAddress.delegatecall(
    abi.encodeWithSignature("targetFunction()"));
  return success;
 }
}
\end{lstlisting}
\centering
\caption{An example of execution status inconsistency. The executor of Solc runs successfully, while the executor of Revive fails.}
\label{delegatecallBugProgram}
\end{minipage}
\begin{minipage}[t]{1\linewidth}
\begin{lstlisting}[language=Rust,
    mathescape, 
    firstnumber=1, 
    escapechar=\%,
    linebackgroundcolor = 
    {\color{verylightgray}\ifnum\value{lstnumber}=4\color{githubpink}
    \else\fi
    }]
fn delegate_call(...) -> ... {
 // ...
 let code_hash = ContractInfoOf::<T>::get(&address)
  .ok_or(Error::<T>::CodeNotFound)
  .map(|c| c.code_hash)?;
 let executable = E::from_storage(code_hash, self.gas_meter_mut())?;
 // ...
}
\end{lstlisting}
\centering
\caption{The source code of the executor in Revive that handles delegatecall}
\label{delegatecallBugReason}
\end{minipage}
\end{figure}

\para{Case Study 2: Deleting an Array Element Causes the Array to Be Deleted in Solang} As shown in Figure~\ref{arrayDeleteBugProgram}, when executing this program, the executor of Solc outputs 2, while the executor of Solang outputs 0. This difference occurs because deleting an array element in Solang removes the entire array, causing its length to be reset to zero.

The inconsistency is detected by \system{} as follows. During the rule selection phase, \system{} successfully selectes the rule \texttt{"insert delete operation to remove array elements"} for mutation. This rule effectively transformed a full-array deletion (Figure~\ref{arrayDeleteOriginalProgram}, line 6) into the deletion of specific array elements (Figure~\ref{arrayDeleteBugProgram}, lines 6–7), thereby triggering the inconsistency.

This inconsistency can lead to severe financial losses. Consider a token contract that uses an array to record users’ token balances. When a user withdraws all their tokens, the contract deletes that user's entry in the array. In Solang, deleting an element in the array causes the entire array to be deleted, potentially removing all users’ balance records. As a result, the balances of other users may be lost, demonstrating a serious risk in financial applications.

\begin{figure}[t]
\begin{minipage}[t]{1\linewidth}
\begin{lstlisting}[language=Solidity,
    mathescape, 
    firstnumber=1, 
    escapechar=\%,
    linebackgroundcolor = 
{\color{verylightgray}\ifnum\value{lstnumber}=6\color{githubgreen}
    \else\fi
    }]
contract C {
    function len() public returns (uint ret) {
        uint[] memory data = new uint[](2);
        data[0] = 234;
        data[1] = 123;
        delete data;
        assembly {
            ret := mload(data)
        }
    }
}

\end{lstlisting}
\centering
\caption{The original version of the program in Figure~\ref{arrayDeleteBugProgram}}
\label{arrayDeleteOriginalProgram}
\end{minipage}
\begin{minipage}[t]{1\linewidth}
\begin{lstlisting}[language=Solidity,
    mathescape, 
    firstnumber=1, 
    escapechar=\%,
    linebackgroundcolor = 
{\color{verylightgray}\ifnum\value{lstnumber}=6\color{githubpink}
    \else\ifnum\value{lstnumber}=7\color{githubpink}
    \else\fi\fi
    }
    ]
contract C {
    function len() public returns (uint ret) {
        uint[] memory data = new uint[](2);
        data[0] = 234;
        data[1] = 123;
        delete data[0];
        delete data[1];
        assembly {
            ret := mload(data)
        }
    }
}
\end{lstlisting}
\centering
\caption{An example of execution output inconsistency. The executor of Solc outputs 2, while the executor of Solang outputs 0.}
\label{arrayDeleteBugProgram}
\end{minipage}
\end{figure}

\begin{tcolorbox}
    \textbf{RQ1:} \system{} has found {\bugs} inconsistencies across six Solidity compilers, {\confirmedBugs} of which have been confirmed, and {\fixedBugs} of which have been fixed due to our Git issues. Notably, the inconsistency related to Revive received a bounty reward from the Polkadot community.
\end{tcolorbox}

\subsection{Comparison with SOTA Fuzzers}
We compare \system{} with the other four Solidity compiler fuzzers (i.e., POLYGLOT, afl-compiler-fuzzer, FUZZOL, and solc\_ossfuzz). The comparison focuses on inconsistency detection capability and the quality of generated programs.

\para{Unique Inconsistencies}
Since other fuzzers cannot detect execution result inconsistencies, we only compare the number of detected compilation result inconsistencies. To avoid the interference of seed programs, we only count novel inconsistencies, meaning those that cannot be triggered before mutation but can be triggered after mutation. Only such novel inconsistencies are considered contributions of the fuzzer. We configure each fuzzer with the complete dataset, then run them separately for 4 days and collect the results.

As shown in Table~\ref{inconsistencyDifferentFuzzers}, \system{} is able to detect more compilation result inconsistencies. Only POLYGLOT detected EMI, and the triggering program is shown in Figure~\ref{constantBugProgram}. This program causes Solc to report an error without a clear message, while Sold and Solang provide clear error messages. Figure~\ref{constantpOriginalProgram} shows the original program. POLYGLOT successfully triggered the Solc error by adding public before constant. \system{} fails to detect this inconsistency because it does not successfully select the specific mutation operation \texttt{"add the visibility of a state variable with public"}. However, \system{} identified many other inconsistencies, which also demonstrates its superiority.

\begin{table}[t]
  \centering
  \caption{Inconsistency detection results of different fuzzers. Before\textbar{}after the slash represents the count of EMI and CSI, respectively.}
  \resizebox{0.48\textwidth}{!}{  
    \begin{tabular}{lcccccc}
      \toprule
      \multicolumn{1}{c}{\multirow{2}[4]{*}{Fuzzers}} & \multicolumn{5}{c}{Compiler} & \multirow{2}[4]{*}{Total} \\
      \cmidrule{2-6}          
      & Revive & Zksolc & Solang & Sold & Solar &  \\
      \midrule
      POLYGLOT & -     & -     & 1\textbar{}2   & 1\textbar{}-   & -\textbar{}1   & 2\textbar{}3 \\
      afl-compiler-fuzzer & -     & -     & -\textbar{}1   & -\textbar{}1   & -\textbar{}1   & -\textbar{}3 \\
      FUZZOL & -     & -     & -     & -     & -     & - \\
      solc\_ossfuzz & -     & -     & -     & -\textbar{}1   & -     & -\textbar{}1 \\
      \textbf{\system{}} & \textbf{-\textbar{}2}   & \textbf{-}     & \textbf{-\textbar{}23}  & \textbf{-\textbar{}4}   & \textbf{-\textbar{}2}   & \textbf{-\textbar{}31} \\
      \bottomrule
    \end{tabular}
  }
  \label{inconsistencyDifferentFuzzers}
\end{table}

\begin{figure}[htbp]
\begin{minipage}[t]{1\linewidth}
\begin{lstlisting}[language=Solidity,
    mathescape, 
    firstnumber=1, 
    escapechar=\%,
    linebackgroundcolor = 
{\color{verylightgray}\ifnum\value{lstnumber}=6\color{githubgreen}
    \else\fi
    }]
interface A {function f() external;}
contract B {function g() public {}}
contract C is B {
    function h() external {}
    bytes4 constant s1 = A.f.selector;
    bytes4 constant s2 = B.g.selector;
    bytes4 constant s3 = this.h.selector;
    bytes4 constant s4 = super.g.selector;
}
\end{lstlisting}
\centering
\caption{The original version of the program in Figure~\ref{constantBugProgram}}
\label{constantpOriginalProgram}
\end{minipage}
\begin{minipage}[t]{1\linewidth}
\begin{lstlisting}[language=Solidity,
    mathescape, 
    firstnumber=1, 
    escapechar=\%,
    linebackgroundcolor = 
    {\color{verylightgray}\ifnum\value{lstnumber}=6\color{githubpink}
    \else\fi
    }]
interface A {function f() external;}
contract B {function g() public {}}
contract C is B {
    function h() external {}
    bytes4 constant s1 = A.f.selector;
    bytes4 public constant s2 = B.g.selector;
    bytes4 constant s3 = this.h.selector;
    bytes4 constant s4 = super.g.selector;
}
\end{lstlisting}
\centering
\caption{An example of error message inconsistency. Solc produces a compilation error without a clear error message, while Sold and Solang provide clear error messages.}
\label{constantBugProgram}
\end{minipage}
\end{figure}

\para{Quality of Generated Program}
Most inconsistency-triggering programs can be successfully compiled by Solc. Cases like the one shown in Figure~\ref{constantBugProgram} are rare, as Solc is already a mature and widely adopted compiler. Inconsistency-triggering programs also exhibit diversity, which can be evaluated through coverage. Therefore, we assess the quality of generated programs using both compilation success rate and code coverage. Since there are differences in the program generation speeds of different fuzzers, directly comparing them would lead to inaccurate quality evaluation. Therefore, we randomly selected 100 programs from the dataset, performed mutations for a period of time, and generated 1,000 new programs. We then evaluated the quality of these 1,000 programs.

\begin{figure}[t]
\centering
\includegraphics[width=3.3in]{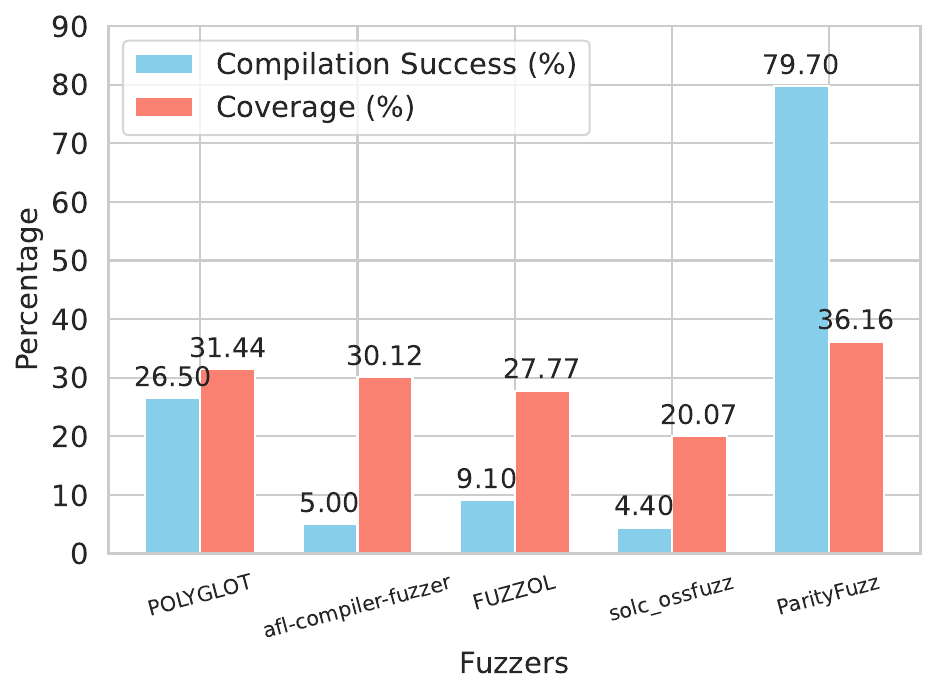}
\vspace{-1ex}
\caption{The quality of 1000 programs generated by different fuzzers.}
\label{programQuality}
\end{figure}

As shown in Figure~\ref{programQuality}, \system{} achieves the highest compilation success rate and coverage rate. \system{} repairs the generated programs, which greatly increases the compilation success rate. \system{} uses two sets of mutation rules to mutate the program, thereby generating complex programs and improving coverage. 

Compared to the second-ranked POLYGLOT, although \system{} achieves only a 15.02\% increase in coverage, it is able to detect more inconsistencies. This is because \system{} performs mutations based on the existing syntactic features of the program, and therefore does not cause drastic changes in coverage. For instance, the original program in Figure~\ref{pushOriginalProgram} has a coverage of 15.71\%, while the mutated version in Figure~\ref{pushBugProgram} has a coverage of 16.92\%. Although the increase in coverage is limited, the mutation successfully produces a program that triggers an inconsistency.

\begin{figure}[htbp]
\begin{minipage}[t]{1\linewidth}
\begin{lstlisting}[language=Solidity]
contract C {
  struct S {
    uint x;
  }
  struct T {
    S s;
    uint y;
  }
  function test() pure public {
    S memory inner = S(43);
    T memory outer = T(inner, 512);
    assert(outer.y == 512);
    assert(outer.s.x == 43);
    assert(outer.s.x == 42);
  }
}
\end{lstlisting}
\centering
\caption{The original version of the program in Figure~\ref{pushBugProgram}}
\label{pushOriginalProgram}
\end{minipage}
\begin{minipage}[t]{1\linewidth}
\begin{lstlisting}[language=Solidity,
    mathescape, 
    firstnumber=1, 
    escapechar=\%,
    linebackgroundcolor = 
    {\color{verylightgray}\ifnum\value{lstnumber}=14\color{githubpink}
    \else\fi
    }]
contract C {
  struct S {
    bytes8 val;
  }
  struct U {
    S[] elems;
    mapping(uint => int) attr;
  }
  function example() public {
    U storage u;
    assembly {
        u.slot := 0
    }
    u.elems.push();
    u.attr[0] = 5;
  }
}
\end{lstlisting}
\centering
\caption{An example of compilation status inconsistency. Compilation error with Solang vs success with Solc.}
\label{pushBugProgram}
\end{minipage}
\end{figure}

\begin{tcolorbox}
    \textbf{RQ2:} Compared to other fuzzers, \system{} can detect more inconsistencies, while also generating programs with higher compilation success rates and greater coverage.
\end{tcolorbox}

\subsection{Ablation Study}. To understand the contributions of each module, we use the following variants of \system{} to conduct ablation studies. 

\begin{itemize}
\item \system{}$_{\text{bor, w/o RL}}$ uses boundary-oriented rules to perform mutation, while eliminating the impact of reinforcement learning fine-tuning.

\item \system{}$_{\text{sor, w/o RL}}$ uses syntax-oriented rules to perform mutation, while eliminating the impact of reinforcement learning fine-tuning.

\item \system{}$_{\text{dr, w/o RL}}$ uses direct rules to perform mutation, while eliminating the impact of reinforcement learning fine-tuning. Direct rules are obtained by using the LLM to directly analyze code blocks that contain boundary conditions, with no extra prompt information(i.e., specific boundary condition code, program feature, and syntax-oriented rules). To be compatible with the selection strategy of \system{}, we classify direct rules using the LLM. This allows us to first select an rule category, and then choose a specific direct rule within that category.

\item \system{}$_{\text{r}}$ uses a purely random strategy to select mutation rules.

\item \system{}$_{\text{w/o RL}}$ eliminates the impact of reinforcement learning fine-tuning.
\end{itemize}

Among these variants, \system{}$_{\text{di, w/o R}}$ is used to investigate whether boundary-oriented rules generated through three steps (shown in Figure~\ref{ruleGenerationProcess}) are more effective than those obtained by directly analyzing the source code in a single step. \system{}$_{\text{sor, w/o RL}}$ and \system{}$_{\text{bor, w/o RL}}$ are designed to study whether syntax-oriented and boundary-oriented rules are beneficial for generating inconsistency-triggering programs. \system{}$_{\text{r}}$ is designed to investigate whether the rule selection strategy of \system{} outperform a purely random strategy. \system{}$_{\text{w/o RL}}$ is designed to evaluate whether LLMs fine-tuned via reinforcement learning can better select mutation rules.

We randomly selected 1,000 programs from the dataset for the ablation study. The five variants and \system{} each performed one round of mutation. Table~\ref{ablation} presents the comparison results between the variants and \system{}. 

\para{Contribution of Rule Generation} As shown in Table~\ref{ablation}, compared to \system{}$_{\text{di, w/o R}}$, \system{}$_{\text{bor, w/o RL}}$ detects more inconsistencies. This indicates that the three-step approach (Figure~\ref{ruleGenerationProcess}) generates more concise and clearer mutation rules, which better guide programs to trigger boundary cases. In contrast, \system{}$_{\text{di, w/o R}}$ combines analysis and generation in a single prompt, often resulting in vague or incomplete rules.

Both \system{}$_{\text{bor, w/o RL}}$ and \system{}$_{\text{sor, w/o RL}}$ are able to detect a large number of inconsistencies, indicating that boundary-oriented rules and syntax-oriented rules are effective. \system{}$_{\text{w/o RL}}$, which combines both sets of mutation rules, is still capable of detecting many inconsistencies, thereby validating the effectiveness of the two sets of mutation rules.

\para{Contribution of Rule Selection} 
Compared to \system{}$_{\text{w/o RL}}$, \system{}$_{\text{r}}$ detects fewer inconsistencies. This is because randomly selected mutation rules may not be applicable to the seed programs. For instance, if a seed program lacks a certain syntactic feature required by a mutation, the resulting program may be invalid.

\para{Contribution of Fine-tuning} Compared to \system{}$_{\text{w/o RL}}$, \system{} detects more inconsistencies. This indicates that the fine-tuned LLM is able to select mutation rules that better match the seed programs.

\begin{table}[t]
  \centering
  \caption{Comparison of inconsistency detection capability with the five variants of \system{}}
    \begin{tabular}{cccc}
    \toprule
    Approach. & Inconsistency  \\
    \midrule
    \system{}$_{\text{bor, w/o RL}}$ & 14  \\
    \system{}$_{\text{sor, w/o RL}}$ & 13  \\
    \system{}$_{\text{dr, w/o RL}}$ & 11  \\
    \system{}$_{\text{r}}$ & 12  \\

    \system{}$_{\text{w/o RL}}$ & 14  \\
    
    \textbf{\system{}} & \textbf{17}  \\
    \bottomrule
    \end{tabular}%
  \label{ablation}%
\end{table}%

\begin{tcolorbox}
    \textbf{RQ3:} Each component of \system{} contributes to detecting more inconsistencies.
\end{tcolorbox}

\section{Discussion}\label{sec:discussion}
\para{Threats to Validity} One threat to validity may stem from slight discrepancies between the execution environment used by \system{} and the actual environments in which blockchains are deployed. To ensure execution efficiency, \system{} prioritizes the use of executors bundled with Solidity compiler projects (e.g., Revive, Solang). These executors can be compiled into executable files to efficiently execute input bytecode. In practical applications, the execution environment is typically a node, which provides a more comprehensive set of functionalities. For example, the executor bundled with Revive does not support precompiled contracts, causing the return value of \texttt{ripemd160} to always be 0, while such functionality is supported by nodes in the blockchain mainnet.

Another threat to validity arises from the fact that some seed programs may trigger inconsistencies without mutation. Therefore, when comparing different fuzzers, we only considered programs where inconsistencies are triggered after mutation. After accounting for the influence of such seed programs, \system{} still detects a significant number of inconsistencies. As a result, the impact of this threat can be considered negligible.

Finally, as \system{} relies on LLM-based program mutation and repair, the nondeterministic nature of LLM outputs could introduce variability in the generated programs. To mitigate this, we fix both the random seed and the temperature of the LLM during mutation and repair, ensuring reproducibility. In practice, multiple runs on the same seed program consistently detect the same cross-compiler inconsistencies, indicating that LLM nondeterminism has a limited impact on our results.

\para{Limitations} \system{} supports various Solidity compilers, demonstrating strong versatility. However, there are two factors that may weaken its generality. First, false positive elimination requires manual intervention. We must manually summarize the Solidity program features that may lead to discrepancies based on the documentation. However, since the relevant documentation is relatively concise, introducing a new compiler does not demand significant human resources. 

Second, some compilers may lack readily available execution environments. For example, the Revive and Solang compilers do not directly provide executors, requiring manual modification of their test programs to wrap them into usable executors. Meanwhile, the Sold compiler relies on a test node as its execution environment, but certain bugs currently prevent it from functioning properly.


\section{Related Work}\label{sec:relateWork} 
Fuzz testing is one of the most effective techniques for uncovering compiler inconsistencies. It typically involves two main steps: program generation and bug detection.

\para{Program Generation} Several tools \cite{wang2023zero,tu2022detecting,ossfuzz} can generate syntax-correct programs from Protobuf specifications \cite{protobuf}. In addition, there are also several mutation-based tools. FUZZILLI \cite{gross2023fuzzilli} and POLYGLOT \cite{chen2021one} mutate the Intermediate Representation of seed programs to generate new programs. While Superion \cite{wang2019superion}, GrayC \cite{even2023grayc}, and FUZZOL \cite{mitropoulos2023syntax} mutate the Abstract Syntax Tree of seed programs. AFL-compiler-fuzzer \cite{groce2022making} decomposes seed programs into small parts and then randomly substitutes them with others to generate new programs. At the same time, LLMs are also widely used for program generation. FuzzGPT \cite{deng2024large} leverages LLMs to generate uncommon input programs for effective fuzzing. WhiteFox \cite{yang2023white} employs LLMs to analyze optimization source code and produce test programs that precisely trigger specific optimizations. Building on the idea of using LLMs for program generation, LLM4CBI \cite{tu2024isolating} further incorporates both data-flow and control-flow analyses to craft more precise prompts for generating targeted programs. Beyond direct program generation, LLMs can also be utilized to synthesize generators or mutators, which in turn are used to produce test programs. MetaMut \cite{ou2024mutators} uses LLMs to create 118 semantic-aware mutators, which can perform diverse mutations on programs. CKGFuzzer \cite{xu2024code} uses LLMs to automatically generate fuzz drivers and perform data flow analysis to produce input seeds, thereby achieving fully automated fuzz testing.

While these approaches are effective for general fuzzing or optimization testing, they are not explicitly designed to trigger cross-compiler inconsistencies. In contrast, \system{} generates fine-grained mutation rules by analyzing the source code of compilers and executors, specifically targeting programs that are likely to expose inconsistencies across compilers.

\para{Differential Analysis} Differential testing is a commonly used approach for detecting inconsistencies. Some tools \cite{wang2023fuzzjit,ossfuzz,bernhard2022jit} detect compiler bugs by comparing outputs generated under different optimization levels within the same compiler, while others \cite{tu2022detecting,ofenbeck2016randir,chen2016coverage} do so by comparing the outputs of different compilers. There are also studies focusing on eliminating false positives during the detection of inconsistencies. Tools such as \cite{sun2016finding,wang2024rustlantis} define various patterns that are likely to cause false positives, and deliberately avoid generating such programs during the program generation phase. FuzzJIT \cite{wang2023fuzzjit} creates a blacklist of disturbing APIs and block their generation during mutation. 

Similar to existing work \cite{tu2022detecting,ofenbeck2016randir,chen2016coverage}, \system{} detects inconsistencies across different compilers or executor. The outputs from compilers can be directly compared, whereas for outputs generated by the executor, \system{} performs a comparison by first decoding the data. In \system{}, various program patterns that may lead to false positives are manually defined, and \system{} eliminates false positives through pattern matching.

\section{Conclusion}\label{sec:conclusion}
This paper proposes \system{}, a cross-compiler differential testing tool to detect inconsistencies among different Solidity compilers. To generate test programs that can lead to inconsistencies, \system{} generates fine-grained mutation rules by analyzing source code (i.e., compiler and execution environment). Then, to select the most suitable mutation rules for seed programs, \system{} first chooses a rule category, and then selects a specific rule within that category. Meanwhile, \system{} uses reinforcement learning to optimize the selection process. Finally, to detect inconsistencies, \system{} compares the results of multiple compilations and executions.

Evaluation results show that \system{} has found {\bugs} inconsistencies among six Solidity compilers. Following our responsible disclosure of the identified inconsistencies, {\fixedBugs} of the newly discovered inconsistencies have been successfully fixed by the community, and the inconsistencies related to Revive received a bounty reward from the Polkadot community. As a next step, we compare \system{} with existing state-of-the-art fuzzers to assess its relative effectiveness. Our experiments show that \system{} is more effective in generating high-quality test programs (up to 18$\times$ improvement in compilation success rate, 1.8$\times$ improvement in coverage) and detecting inconsistencies than state-of-the-art fuzzers (31$\times$ more inconsistencies).



\printcredits

\section*{Declaration of competing interest}
The authors declare that they have no known competing financial interests or personal relationships that could have appeared to influence the work reported in this paper.

\section*{Acknowledgments}
This work is supported by NSFC/RGC Collaborative Research (62461160332), Natural Science Foundation of China (62032025, 624B2139) and Guangdong Zhujiang Talent Program (2023QN10X561).

\section*{Data availability}
The source code and experimental results are publicly available at: \url{https://zenodo.org/records/19944888}



\bibliographystyle{cas-model2-names}

\bibliography{main}





\end{document}